\documentclass[aps,prl,twocolumn, superscriptaddress]{revtex4-2}
\pdfoutput=1
\usepackage{graphicx}
\usepackage{grffile} 
\usepackage{dcolumn}
\usepackage{bm,amssymb,amsmath}
\usepackage[pdftex]{hyperref}
\hypersetup{colorlinks=true,citecolor=blue,linkcolor=red,urlcolor=blue}
\usepackage[all]{hypcap}
\usepackage{color}
\definecolor{red}{rgb}{0.8, 0.0, 0.0}
\definecolor{blue}{rgb}{0.06, 0.2, 0.65}
\definecolor{green}{rgb}{0,0.6,0}

\begin{document}
\title{Avalanches in CuZrAl metallic glasses}
\author{Tero Mäkinen}%
\email{tero.j.makinen@aalto.fi}
\affiliation{Aalto University, Department of Applied Physics, P.~O. Box 15600, 00076 Aalto, Espoo, Finland}
\author{Anshul D. S. Parmar}
\email{Anshul.Parmar@ncbj.gov.pl}
\affiliation{NOMATEN Centre of Excellence, National Center for Nuclear Research, ul. A. Soltana 7, 05-400 Swierk/Otwock, Poland}
\author{Silvia Bonfanti}
\affiliation{NOMATEN Centre of Excellence, National Center for Nuclear Research, ul. A. Soltana 7, 05-400 Swierk/Otwock, Poland}
\author{Mikko J. Alava}
\affiliation{Aalto University, Department of Applied Physics, P.~O. Box 15600, 00076 Aalto, Espoo, Finland}
\affiliation{NOMATEN Centre of Excellence, National Center for Nuclear Research, ul. A. Soltana 7, 05-400 Swierk/Otwock, Poland}

\begin{abstract}
Metallic glasses have mechanical properties, which exhibit avalanches in the disguise of stress drops. We study these phenomena in a classical metallic glass system CuZrAl by athermal quasistatic shear and varying the element concentrations and for pure CuZr 50/50 case the cooling rate. The resulting mechanical properties are close to the behaviour found experimentally. At small strains, the pristine systems are akin to other glassy systems with a so-called gap distribution with a small positive exponent. Critical avalanching behaviour is found only approaching the yield point. The post-yield stress drops are universal, and the gap distribution becomes flat.
\end{abstract}

\maketitle

\section{Introduction}

Despite the no-long-range structural order, metallic glasses~(MGs) exhibit elasticity to the external shear and will flow if sufficient shear stress is applied. The plasticity can be conceived as consisting of elementary rearrangements, that are localized in the so-called shear transformation zones~(STZs), but display long-range elastic interactions. Below a critical strain, the events are localized~\cite{argon1979plastic,falk1998dynamics,tanguy2006plastic,li2013shear} and result in crackling noise -like stress drops. The piecewise elastic response is intersected with these plastic stress drops and as one approaches the yield point, the plastic events manifest as nonlocalized avalanches, and finally, the system reaches a well-defined steady state. 

As an elementary observation, the statistics of such plastic excitations correlate with the microscopic description. The probability distribution function $p(\Delta \gamma; \gamma)$ in the strain interval over accumulated strain~$\gamma$, is defined by the additional shear strain~$\Delta \gamma$ required to trigger an avalanche at the strain~$\gamma$~\cite{lemaitre2007plastic,karmakar2010statistical,hentschel2015stochastic}. This is conceptually useful also in order to make a connection to depinning transition theory \cite{lin2014scaling}. 
The empirical pseudo-gap, i.e.~$\lim_{\Delta \gamma \rightarrow 0} p(\Delta\gamma; \gamma=0^{+})\propto \Delta \gamma ^\theta$ is characterized by the power-law behavior of the~$\Delta \gamma$ distribution at small values. The exponent $\theta$ does not claim any universality and is found to be dimension and glass preparation protocol dependent~\cite{karmakar2010statistical,lin2014scaling,lin2015criticality}. 
However, as the yield is approached from below, initially, the exponent rapidly drops and in the steady state becomes zero beyond the yield strain~\cite{hentschel2015stochastic,lin2015criticality,lin2016mean}. 
In the simplistic picture, in response to the deformation, the plastic rearrangements and their spatial and temporal correlations can be considered in the context of the pinning to depinning transition~\cite{baret2002extremal}, making a depinning event the epicentre of an avalanche. The role of the critical threshold in the depinning against the flow and the growth of the local strain can explicitly be considered for the macroscopic yield stress and the shear band~\cite{baret2002extremal,nicolas2018deformation,rosso2022avalanches}. 

The studies of model glass systems by molecular dynamics extend over a long period of time with the main questions being asked concentrating on the critical exponents of the yield transition and the effects of inertia and temperature~\cite{PhysRevE.88.062206,karmakar2010statistical}. These naturally have been based on the~STZ and depinning analogies and later this work has expanded to consider the effect of sample preparation or cooling. The current understanding points towards a critical point in disorder, separating systems with a finite stress drop (``good glasses'') from those with no drop (``bad glasses'')~\cite{OzawaPNAS}. While this is indicated by mean-field arguments the properties of the steady-state beyond the yield stress and the approach to that are an open question in terms of universality, apart from the generic idea of a gap distribution, applied both below and above the yield stress. 

Recently some work has been done on more realistic (metallic) glass models, including the CuZr system, including aspects like the geometric details of the yield events as plastic strain develops~\cite{sun2010plasticity,niiyama2019structural,cui2022anisotropic,MartensEPJ} with partial evidence for the avalanche statistics in general. One may confront these with generic glass criticality~\cite{OzawaPNAS} and generally a stress drop exponent for the distribution~$p(\Delta \tau$) of stress drops~$\Delta \tau$ emerges with a value of about~$1.2 \dots 1.4$. As usual, the distribution is via its cutoff~$\Delta \tau_0$ coupled to the value of the mean drop at a given strain, $\langle \Delta \tau \rangle$ as noted by Shang~\textit{et al.}~\cite{doi:10.1073/pnas.1915070117}. Interestingly in the Lennard-Jones system studied there power-law like activity was found for small~$\gamma$.

The CuZr(Al) system is the ubiquituous metallic glass for its good glass forming ability and the possibility of controlling its properties again by cooling but also by the composition. In particular it was recognized some time ago~\cite{inoue2002formation} that aluminum brings beneficial effects as it seems by forming local structural motifs to improve mechanical properties. Thus it is of interest to compare this key glass family \textit{in silico} to expectations from glass theory and to experimental observations.

For instance, the experimental work of Das~\textit{et~al.}~\cite{das2005work} shows that the composition Cu$_{47.5}$Zr$_{47.5}$Al$_5$ has superior mechanical properties, such as higher strength, improved ductility, and work-hardenability, compared to the binary Cu$_{50}$Zr$_{50}$, prepared by arc melting. 
Same optimal 5~\%~Al was found by Yu~\&~Bai~\cite{yu2008poisson} when considering maximal plastic strain and also the Poisson ratio.
Furthermore, it has been shown by Pauly~\textit{et~al.}~\cite{pauly2010transformation} that addition of Al increases tensile failure stress and elastic modulus, but decreases the plastic strain.
In nanoindentation experiments Poltronieri~\textit{et~al.}~\cite{POLTRONIERI2023119226} found an increase in shear and elastic moduli as well as hardness with increasing Al content~(up to 12~\%). Similarly, Cheung~\&~Shek~\cite{cheung2007thermal} in their nanoindentation experiments found increase in hardness and elastic modulus with increasing Al content~(up to 10~\%). Additionally they saw a decrease in creep displacement.\\

The structure of this paper is as follows. In the next section we briefly go through the methodology needed to study the shear response of MGs in silico. The results part is divided roughly into three themes: the generic mechanical properties and their dependence on cooling (pure CuZr) and composition, the small-strain regime before yielding, and the statistics of yielding in the steady-state after that. Finally, we finish with a summary.

\begin{table}[tb!]
     \centering
     \begin{tabular}{|l|c|c|c|c|c|}
      \hline
      Composition & $N$ & $\dot{T}$ & $G$ & $\tau_{\rm max}$ & $\gamma_{\rm max}$ \\
      & & [K/s] & [GPa] & [GPa] & \\
      \hline  
       Cu${}_{0.50}$Zr${}_{0.50}$ & 6000 & $10^{11}$ & 24.74 & 1.811 & 0.1095 \\ 
       Cu${}_{0.50}$Zr${}_{0.50}$ & 6000 & $10^{12}$ & 23.74 & 1.760 & 0.1035 \\
       Cu${}_{0.50}$Zr${}_{0.50}$ & 6000 & $10^{13}$ & 22.90 & 1.537 & 0.1035 \\ 
       Cu${}_{0.50}$Zr${}_{0.50}$ & 6000 & $10^{14}$ & 20.47 & 1.049 & 0.0945 \\ 
       Cu${}_{0.50}$Zr${}_{0.50}$ & 6000 & $10^{15}$ & 16.95 & 0.845 & 0.9025 \\ 
       Cu${}_{0.30}$Zr${}_{0.70}$ & 6000 & $10^{12}$ & 19.00 & 0.987 & 0.0945 \\
       Cu${}_{0.70}$Zr${}_{0.30}$ & 6000 & $10^{12}$ & 24.84 & 1.580 & 0.1065 \\
       Cu${}_{0.47}$Zr${}_{0.47}$Al${}_{0.06}$ & 6000 & $10^{12}$ & 24.00 & 1.743 & 0.1005 \\
       Cu${}_{0.45}$Zr${}_{0.45}$Al${}_{0.10}$ & 6000 & $10^{12}$ & 24.93 & 1.736 & 0.1045 \\
       Cu${}_{0.40}$Zr${}_{0.40}$Al${}_{0.20}$ & 6000 & $10^{12}$ & 26.59 & 1.799 & 0.1095 \\
       Cu${}_{0.37}$Zr${}_{0.37}$Al${}_{0.26}$ & 6000 & $10^{12}$ & 28.35 & 1.845 & 0.1115 \\
       Cu${}_{0.50}$Zr${}_{0.50}$ & 3000 & $10^{12}$ & 23.43 & 1.667 & 0.1025 \\
       Cu${}_{0.50}$Zr${}_{0.50}$ & 12000 & $10^{12}$ & 23.17 & 1.798 & 0.1025 \\
       Cu${}_{0.50}$Zr${}_{0.50}$ & 24000 & $10^{12}$ & 23.96 & 1.834 & 0.1045 \\
          \hline
     \end{tabular}
     \caption{Averaged mechanical parameters for the compositions used (systems of size 6000 shown in Figure~\ref{figure1}).
     The columns are system size~$N$, the cooling rate~$\dot{T}$, the shear modulus~$G$, the maximum shear stress~$\tau_{\rm max}$, and the strain corresponding to the maximum stress~$\gamma_{\rm max}$.}
     \label{tab:mechanical_prop}
 \end{table}

\section{Methods}

\textit{Simulation methods ---}
Metallic glass samples are simulated using molecular dynamics methods implemented in~LAMMPS~\cite{lammps}, where the interactions are modelled through the embedded atom method~(EAM) as developed by~Ref.~\cite{cheng2009atomic}. 
The simulations are performed in a cubic box with periodic boundary conditions in three dimensions. 
We vary the element concentrations, system sizes~$N$, and cooling rates~$\dot{T}$ as reported in~Tbl.~\ref{tab:mechanical_prop}.\\

\textit{Hybrid molecular dynamics + Monte Carlo~(MD+MC) algorithm ---} To generate metallic glasses we employ a hybrid molecular dynamics + Monte Carlo (MD+MC) scheme under the variance-constrained semi-grand canonical ensemble~(VC-SGC)~\cite{sadigh2012scalable}. This~VC-SGC~MC scheme allows exploring the configurational degrees of freedom by randomly selecting an atom and attempting to change its type, while also calculating the corresponding energy and \textit{concentration} changes. It allows to target specific concentration ranges while maintaining a fixed total number of particles and volume. Acceptance of these transmutations follows the Metropolis criterion, ensuring the preservation of detailed balance. 
On the other hand, the relaxation processes are accounted for by the~MD integration steps. To maintain the desired concentration within the system~\cite{sadigh2012scalable}, we set the variance parameter~$\kappa=10^3$. The differences in chemical potential relative to~Zr using hybrid~MD+MC simulations under the semi-grand canonical ensemble at a temperature of~2000~K and the specific set of parameters that minimize the composition errors in relation to the desired concentration can be found in~Ref.~\cite{alvarez2023simulated}.
The hybrid scheme is also used for~CuZr by~Ref.~\cite{zhang2022shear}.\\

\textit{Quenching Protocol ---}
The glass state is obtained through quenching in the isobaric-isothermal ensemble~($NpT$) from the liquid at~2000~K to~300~K using a cooling rate reported in~Tbl.~\ref{tab:mechanical_prop} and a time step~$\Delta t =1$~fs.
For the~MD+MC scheme, every~20~MD steps, a~MC cycle consisting of~$N/4$ attempts is performed.
The cooling process is performed by integrating the Nos\'e-Hoover equations with damping parameters~$\tau_T=2$~fs and~$\tau_p=5$~ps for the thermostat and barostat. All results are obtained keeping the external pressure~$P=0$.\\

\begin{figure}[tb!]
	\centering
	\includegraphics[width=\columnwidth]{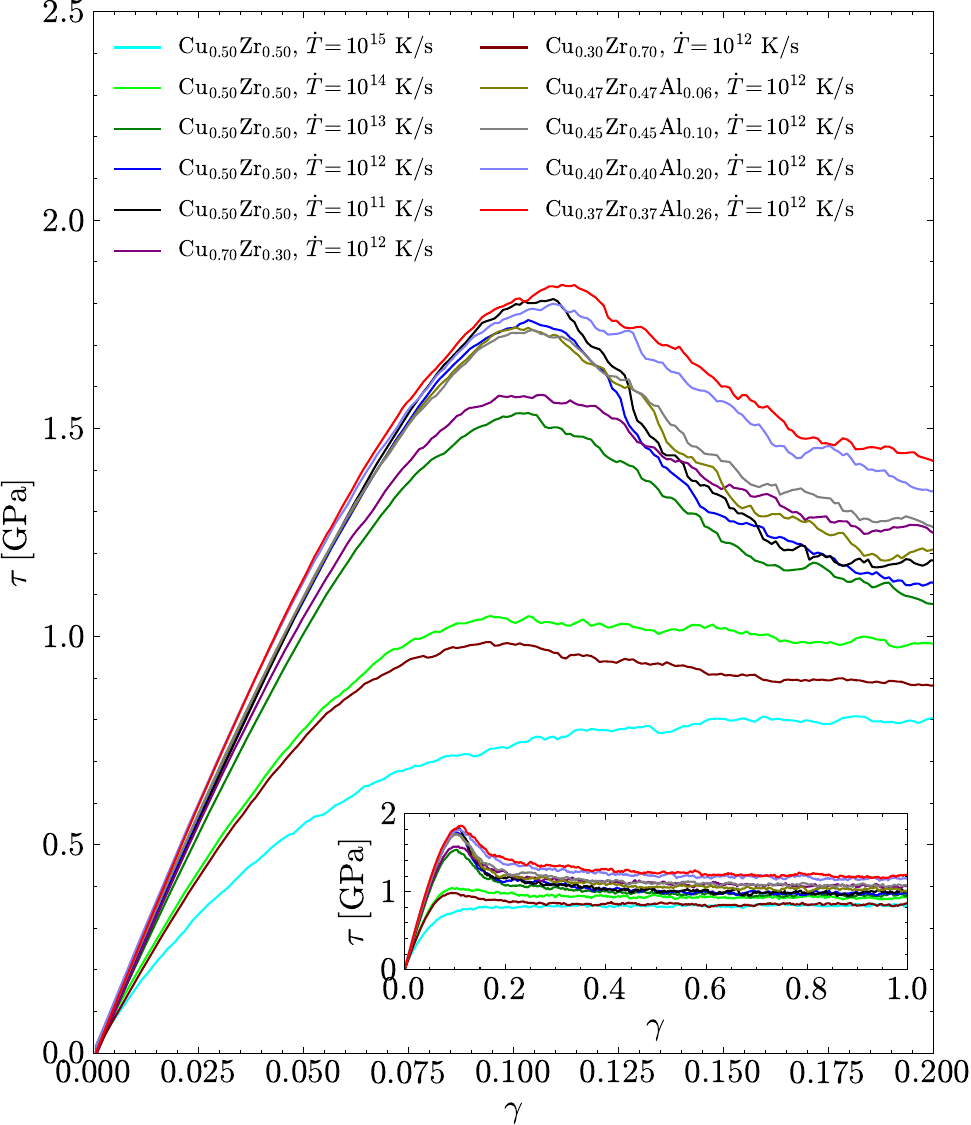}
	\caption{
    The averaged stress-strain curves for various compositions and cooling rates~$\dot{T}$ (for~$N = 6000$).
    The inset shows a zoomed-out view of the same data up to $\gamma = 1$, showing the steady-state behavior.
    }
	\label{figure1}
\end{figure}

\textit{Athermal quasistatic shear ---}
The shear simulations are performed for the samples cooled to~$T=300$~K. The simulation box is incrementally sheared along the~$x$-direction with respect to the~$y$-direction by~$\delta\gamma = 10^{-4}$, and at each strain increment,~MD simulations are performed for a duration of~1~ps. Throughout the shearing process, we record the stress-strain response, which provides a detailed stress-strain curve for analysis~(see~Fig.~\ref{figure1}).
From the stress-strain curves we determine the shear modulus~$G$, by averaging the positive slopes of the stress-strain curve for the first~1~\% of strain in each configuration.
The maximum stress~$\tau_{\rm max}$ and the corresponding strain~$\gamma_{\rm y}$~(shown in Tbl.~\ref{tab:mechanical_prop}) are then determined from the stress-strain curve averaged over all the configurations, by taking the point corresponding to the maximum stress.

The stress drops are determined by looking at the periods when the stress~$\tau$ is decreasing. The drop size~$\Delta \tau$ is the difference between the the point where the stress started to decrease and the point when it starts to increase again. Drops smaller than~$10^{-4}$~GPa are discarded. The strain increments~$\Delta \gamma$ are then the lengths of the periods between the end of a previous stress drop and the start of the subsequent drop.\\

\textit{Maximum likelihood fitting ---}
The distributions observed are fitted using the maximum likelihood method~\cite{baro2012analysis}. For a dataset $x_i$ ($i$ ranging from 1~to~$N_x$) and a probability density function $p(x; \phi_j)$, where $\phi_j$ are the parameters for the distribution,
one computes the parameters that maximize the likelihood function $\mathcal{L}(\phi_j) = \prod_{i=1}^{N_x} p(x_i; \phi_j)$, which are the optimal parameters.

For the case of the Weibull distribution we emphasize the pseudo-gap part of the distribution, and initially do the fitting only from zero to a value $\Delta \gamma = 10^{-3}$ corresponding roughly to the mode of the distribution. Fixing this estimate for the pseudo-gap exponent~$\theta$, we then fit the full distribution, using the whole dataset.

\section{Results}

\subsection{Generic behavior}

Fig.~\ref{figure1} and Tbl.~\ref{tab:mechanical_prop} show the main features of different compositions and/or cooling rates as regards the typical mechanical properties under shear. The curves of Fig.~\ref{figure1} (averages of 100 runs each) split roughly into two groups: ones with a pronounced stress peak (which makes it easy to define a yield stress) and ones without. The latter consists of a 30/70 CuZr mixture and two very fast-cooled 50/50 cases. 
The highest yield stresses are achieved with high Al contents, but the sharpest peaks with the slowest cooling rates.
The Tbl.~\ref{tab:mechanical_prop} shows the resulting mechanical characteristics with clear trends for the elastic modulus and peak/yield stress with the cooling rate and likewise a weaker but still clear trend with the addition of Al to the equiatomic CuZr composition. The results of Tbl.~\ref{tab:mechanical_prop} for different~$N$ give some confidence limits to the values quoted, as there is a slight trend with $N$. The effect of Al is in agreement with the experimental picture (e.g. \cite{POLTRONIERI2023119226}) in that the shear modulus shows a clear but non-drastic increawse with Al. Note that the change in the yield stress is in relative terms larger. After the peak, the systems enter in a flow state with the flow stress following the same order from case to case~(see inset of~Fig.~\ref{figure1}).

\begin{figure}[bt!]
    \centering
    \includegraphics[scale=1]{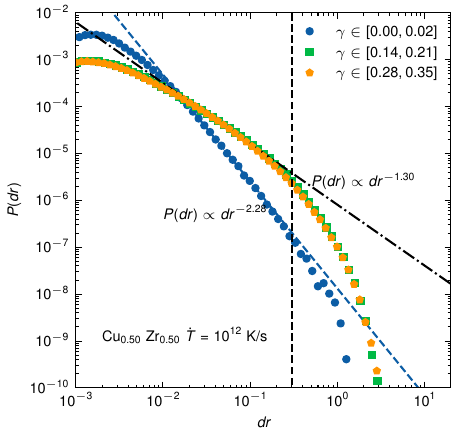}
    \caption{Distributions of the particle's non-affine displacements at the plastic events shown in log-log scale for various strain intervals for a system of Cu$_{50}$Zr$_{50}$ with the cooling rate $\dot{T} = 10^{12}$~K/s for 6000 particles. The particles are considered plastic/active if the non-affine displacement corresponds to the tails of the distribution, marked as $dr \ge 0.3 $~{\AA}~\cite{schroder2000crossover}.}
    \label{Pdr}
\end{figure}

\begin{figure}[t!]
    \centering
    \includegraphics[scale=1]{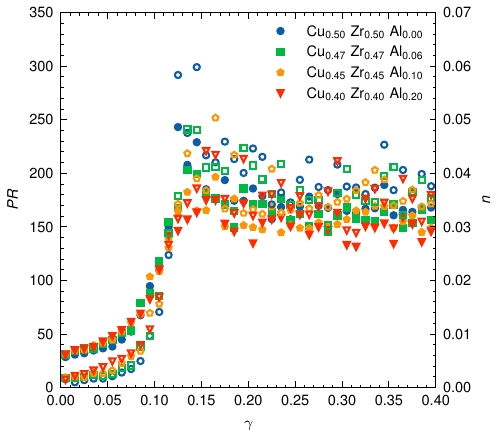}
    \caption{For a plastic event in various compositions, the participation ratio ($PR$, filled symbols) and the fraction of particles in plasticity ($n$, open symbols) for the various concentrations of Al remains the same across the yield.}
    \label{PR_Alxx}
\end{figure}

\subsection{Development of plastic deformation}

Next, we consider the effects of cooling and composition with the microscopic picture of the plastic deformation. In the presence of an avalanche, the system can be viewed as the plastically deformed particles as the core and the rest of the system undergoing an elastic deformation in response to the stresses created by the imposed strain. 
To this end, we first estimate the particles participating in a plastic deformation by considering the non-affine displacement of particles. It is observed that the distribution of single-particle displacements~$p(dr)$ displays a power-law elastic profile with an exponential tail at the cutoff. The deviations from the power law account for the particles ``participating" in the plastic deformations~\cite{schroder2000crossover}.
Fig.~\ref{Pdr} shows the single-particle displacements for the plastic events in various windows of strain. The exponent of the power law, however, is found to be sensitive to the window of the strain~\cite{leishangthem2017yielding}. Particles are considered ``plastic" if they are displaced by more than 0.3~\r{A}. With this cutoff in the particle displacements, we wish to
include particles that take part in the plastic deformation, and exclude particles undergoing elastic response. 

The microscopic nature of the plastic deformation is defined by two different approaches. Firstly, we observe the fraction of the particles involved in the plastic response at the avalanche, i.e.~$n=N_{pl}/N$, where~$N_{pl}$ is the number of particles displaced by more than the cutoff~0.3~\r{A}. For the second approach, we study the extent of localization of particle rearrangements resulting from the transitions between basins at the avalanches. We measure the number of involved particles in the basin's transition to the plastic event by the participation ratio~(PR), defined as
\begin{equation}
    PR = \frac{\left[ \sum_{i=1,N} {d{r}^2_{i}} \right]^2}{\sum_{i=1,N} d{{r}^4_i}}
    \label{eqnPR}
\end{equation}
where $d r_{i}$ is the displacement of particle `$i$' between basins at the plastic event. In the plastic event,
the~PR would yield the total mobile particles involved in the transition. 

\begin{figure}[t!]
    \centering
    \includegraphics[scale=1]{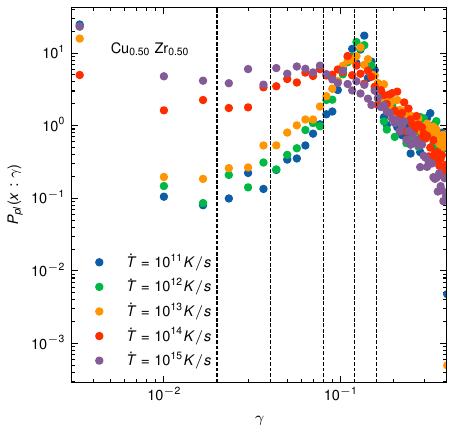}
    \caption{ The distribution of the particle undergoing the first yield event. In better annealed glasses more particles participate around the yield, which is coherent with the conventional ductile-to-brittle response. The dashed lines indicate the strains used in the snapshots in Fig.~\ref{figurevisu}.}
    \label{figurefp}
\end{figure}

Fig.~\ref{PR_Alxx} shows the fraction of plastic particles and the parameter-free description of the avalanche by the participation ratio for the various concentrations of Al and the strain across the tentative yield point. The introduction of the various concentrations of Al has little effect to contribute to the deformation and does not affect the nature of the avalanches at the microscopic level. 

\begin{figure*}[t!]
    \centering
    \includegraphics[scale=0.65]{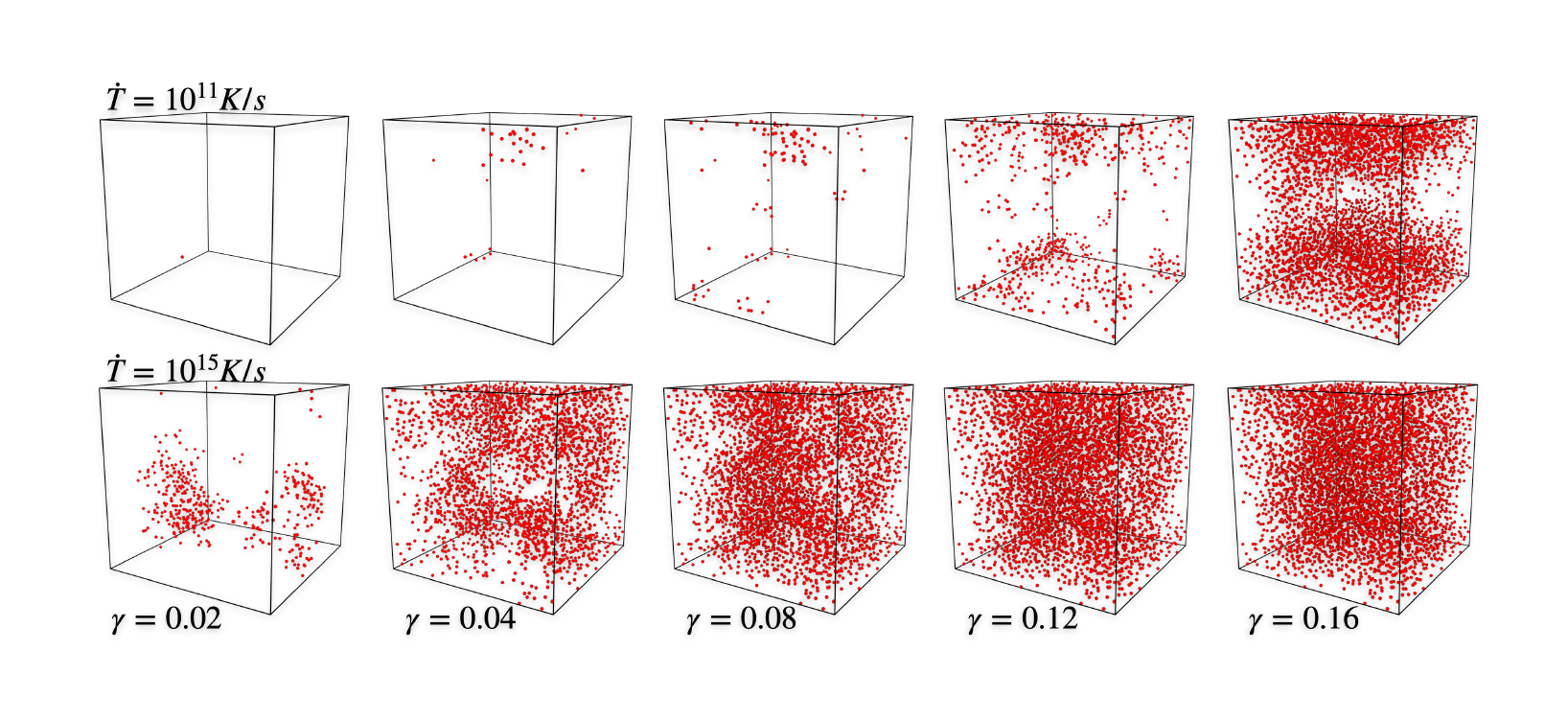}
    \caption{The spatial profile of the particles undergone the first plastic event. Compared with the poor glass, the well-relaxed glass shows localised plastic events, which act as the nucleation points for the additional plastic events.}
    \label{figurevisu}
\end{figure*}

Once established that the nature of avalanches is independent of the compositions, we further focus on the cooling rates dependent on the avalanches for the~CuZr system. To discover the role of cooling in the local stability of the particles against the mechanical deformations, we record the strain for each particle for the first plastic event. Fig.~\ref{figurefp} shows the distribution of the particle undergoing the first plastic event for the various cooling rates for the CuZr glass. 
The fraction of particles undergoing the first avalanche for the~$\gamma \to 0$, can be found consistent with the marginal stability in the glasses~\cite{hentschel2015stochastic}. 
In better-annealed glasses with high stability, a larger fraction of particles undergo the first plastic event in the narrow strain window near the yield. This is consistent with the increasing brittle-like response as the yield approaches. Meanwhile, the yield is gradual for poor glasses, and particles undergo the first plastic deformation across a wide range of strain values.

The question that naturally arises is ``{\it how are the plastic events spatially distributed?}"
Fig.~\ref{figurevisu} shows the spatial arrangement of the particles undergoing the first plastic event for a range of strain and cooling rates, incising better and poorly annealed glasses. For the poor glass, the first plasticity for particles is observed to be strain-independent, highlighting the presence of numerous soft zones and underlying complex elastic interactions. Whereas, for the well-annealed glasses the particles undergoing the first plastic events are somewhat correlated and approach maximal around the yield, manifesting into the shear band. The overall correlation between the plastic events and cooling effect needs further attention, which we wish to address in the subsequent work with larger samples and cooling rates.  

\begin{figure}[bth!]
	\centering
	\includegraphics[width=\columnwidth]{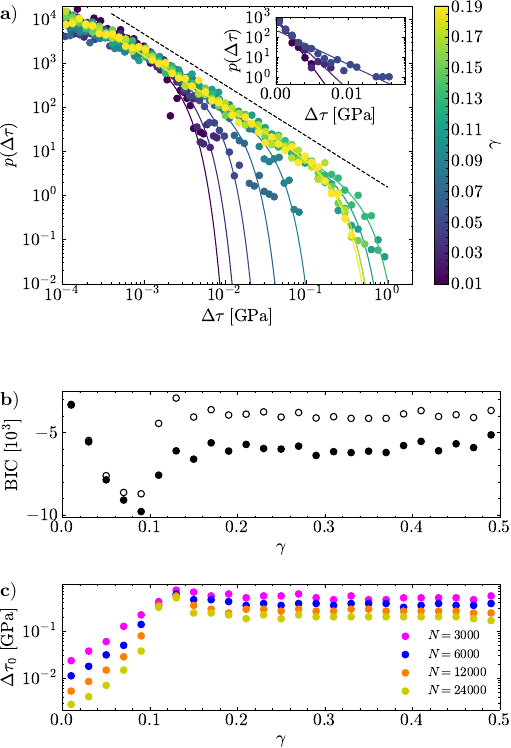}
	\caption{
    \textbf{a})~The stress drop distribution~(for the Cu${}_{0.50}$Zr${}_{0.50}$, $\dot{T} = 10^{12}$~K/s,~$N=24000$ system) for various strain bins indicated by the plot color. The lines are fit to Eq.~\ref{eq:stress_drop_pl} and the dashed black line shows the pure power-law behaviour $p(\Delta \tau) \propto \Delta \tau^{-1.16}$. The inset shows an exponential fit for the three first strain bins, showing the lack of power-law scaling.
    \textbf{b})~The BIC~(Eq.~\ref{eq:bic}) as a function of strain for the power law with Gaussian cutoff (solid dots) and the half-Gaussian distribution (hollow dots).
    \textbf{c})~The cutoff scale of Eq.~\ref{eq:stress_drop_pl} as a function of strain for different system sizes.
	\label{figure2}}
\end{figure}

\begin{figure}[bth!]
	\centering
	\includegraphics[width=\columnwidth]{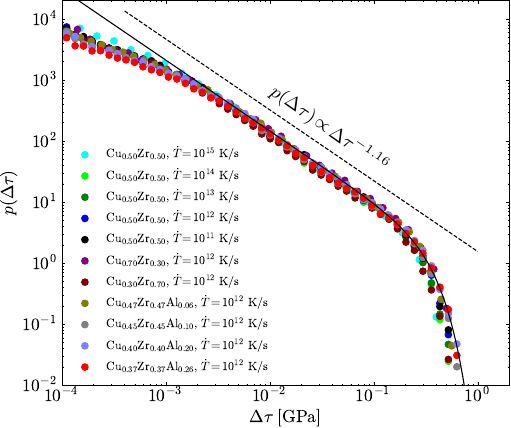}
	\caption{The steady-state stress drop distribution for different compositions and cooling rates~$\dot{T}$ (for $N=6000$). The solid black line corresponds to a fit to Eq.~\ref{eq:stress_drop_pl} which gives a universal result with cutoff $\Delta \tau_0 = 0.34$~GPa. The dashed line gives a reference power-law with the exponent $\beta = 1.16$.}
	\label{figure3}
\end{figure}

\subsection{Stress drops}
As discussed in the Introduction the crackling noise in the stress-strain curve is defined with the drops of stress~$\Delta \tau$ and the intervals between those drops~$\Delta \gamma$, giving rise to the gap distribution~$p(\Delta \gamma)$.

Looking at the stress drops~$\Delta \tau$ in different strain windows shows (see Fig.~\ref{figure2}a) that they follow a power-law distribution with a Gaussian cutoff
\begin{equation} \label{eq:stress_drop_pl}
    p(\Delta \tau) \propto \Delta \tau^{-\beta} \exp\left[ - \left( \frac{\Delta \tau}{\Delta \tau_0} \right)^2 \right]
\end{equation}
where $\beta$ is an exponent and $\Delta \tau_0$ the cutoff scale increasing with strain. Fig.~\ref{figure2}a shows the maximum likelihood~\cite{baro2012analysis} fits to this distribution where over two decades of power-law scaling with an exponent of 1.16 can be observed for strain intervals after the yield point. 

Even though the distribution of Eq.~\ref{eq:stress_drop_pl} fits the data well for all the strain bins, the power-law region is extremely small for small strains.
This means that the distributions can be fitted using a light-tailed distribution, as illustrated in the inset of Fig.~\ref{figure2}a where stress drop distributions for the first three strain bins are plotted using a semilog scale along with an exponential fit.
To quantify this we compute the Bayesian Information Criterion~\cite{leonard2001bayesian, makinen2023history}
\begin{equation} \label{eq:bic}
    \mathrm{BIC} = - 2 \ln \mathcal{L} + n_{\rm p} \ln N_{\Delta \tau}
\end{equation}
where $\mathcal{L}$ is the likelihood function, $N_{\Delta \tau}$~the number of stress drop datapoints, and $n_{\rm p}$~the number of parameters in the chosen distribution. The lower the BIC is, the better the model fits the data. We compute the BIC for a power-law distribution with a Gaussian cutoff (Eq.~\ref{eq:stress_drop_pl} and~$n_{\rm p} = 2$) and for a half-Gaussian distribution (where~$n_{\rm p} = 1$ and $p(\Delta \tau) \propto e^{-(\Delta \tau / \Delta \tau_0)^2}$).
From~Fig.~\ref{figure2}b one can see that both distributions perform roughly equally before the yield point, signifying a lack of a power-law scaling region below this point.
After the yield point, the power-law fits the data significantly better.

In the steady state regime (here $\gamma > 0.50$) the stress drop distributions show completely universal behavior (see Fig.~\ref{figure3}) for all the cooling rates and compositions. The distribution follows Eq.~\ref{eq:stress_drop_pl} with the same exponent~$\beta=1.16$ and an universal cutoff value of $\Delta \tau_0 = 0.34$~GPa.\\

The behavior of the cutoff scale~$\Delta \tau_0$ (see Fig.~\ref{figure2}c) resembles the behavior seen in Figs.~\ref{PR_Alxx}. When approaching yielding from below, the cutoff scale increases rapidly, reaching a maximum around yielding. The value then slightly decreases to a steady-state value, which is achieved already at around 15~\% strain. 

\begin{figure}[tb!]
	\centering
	\includegraphics[width=\columnwidth]{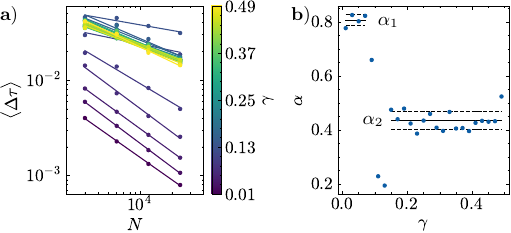}
	\caption{
    \textbf{a})~The average stress drop as a function of the system size for different strain bins is illustrated by the plot colour. The lines correspond to power-law fits $\langle \Delta \tau \rangle \propto N^{-\alpha}$.
    \textbf{b})~The evolution of the exponent~$\alpha$ with the strain. The black lines illustrate the exponent (solid line) and the standard deviation (dashed lines) for the pre-yield ($\alpha_1 = 0.81 \pm 0.02$) and steady-state regimes ($\alpha_2 = 0.44 \pm 0.03$).}
	\label{figure6}
\end{figure}

The behavior of the stress drop cutoff with the system size~(Fig.~\ref{figure2}c) shows a power-law-like scaling below the yield point, roughly system-size independent behavior at yielding, and power-law-like scaling again after yielding.

Looking at the scaling of the average stress drop with the system size (see Fig.~\ref{figure6}a), one can see fairly robust power-law scaling $\langle \Delta \tau \rangle \propto N^{-\alpha}$. However, the exponent~$\alpha$ changes with strain.
Two distinct power-law regimes can be recognised (see Fig.~\ref{figure6}b). At low strains, below the yield point, the average stress drop size scales with an exponent~$\alpha_1 = 0.81 \pm 0.02$. At high strains, above the yield point and close to the steady-state, the exponent value of $\alpha_2 = 0.44 \pm 0.03$ can be found.
Around the yield point, the scaling vanishes as the exponent~$\alpha$ goes close to zero.
For the distribution of Eq.~\ref{eq:stress_drop_pl} the average value scales approximately as
$\langle \Delta \tau \rangle \propto \Delta \tau_0^{1-\beta}$ which is clearly seen in the similar behavior of Figs.~\ref{figure2}c and \ref{figure6}.
When one accounts for the difference in definitions of the average avalanche size (multiplication by a factor~$N$), these exponent values roughly match the results obtained in previous studies~\cite{lemaitre2007plastic, hentschel2015stochastic, karmakar2010statistical, doi:10.1073/pnas.1915070117}.

\begin{figure}[thp]
	\centering
	\includegraphics[width=\columnwidth]{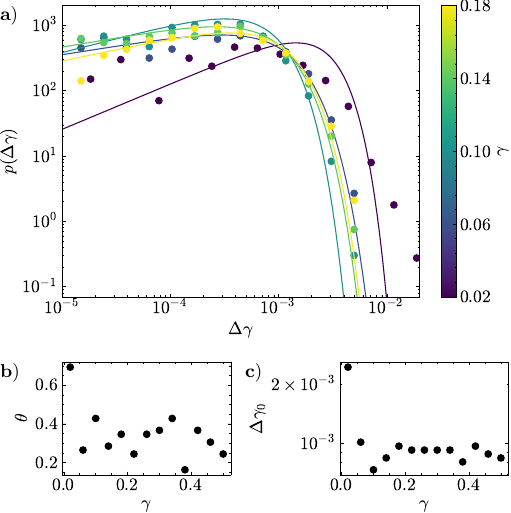}
	\caption{
    \textbf{a})~The gap distribution or the strain interval distribution (for Cu${}_{0.50}$Zr${}_{0.50}$,~$\dot{T} = 10^{11}$~K/s,~$N=24000$) for various strain bins indicated by the plot color.
    The lines are fits to Eq.~\ref{eq:gap_dist}.
    \textbf{b})~The exponent of Eq.~\ref{eq:gap_dist} as a function of the strain.
    \textbf{c})~The cutoff scale of Eq.~\ref{eq:gap_dist} as a function of the strain.}
	\label{figure4}
\end{figure}

\subsection{Gap distribution}

To characterize the behavior of the pseudogap, we have fitted the gap distribution using the Weibull distribution~(see Fig.~\ref{figure4}a)
\begin{equation} \label{eq:gap_dist}
    p(\Delta \gamma) \propto \Delta \gamma^{\theta} \exp\left[ - \left( \frac{\Delta \gamma}{\Delta \gamma_0} \right)^{1+\theta} \right]
\end{equation}
where $\theta$ corresponds to the gap exponent and $\Delta \gamma_0$ to the cutoff scale.

The values of the gap exponent start (at low strains) close to $\theta = 0.7$ and rapidly decrease to values around $\theta = 0.3 \pm 0.1$ (see Fig.~\ref{figure4}b).
The cutoff scale $\Delta \gamma_0$ also starts from a high value, decreases to a minimun value around yielding, and then increases to a steady-state value (see Fig.~\ref{figure4}c), mirroring the behavior of the stress drop cutoff~$\Delta \tau_0$.

It is noteworthy that the tail of the distribution in the first strain bin does not follow the Weibull distribution very well. Instead, it shows a much wider tail, extending to half of a decade higher values than the distributions of the other strain bins.\\

\begin{figure}[thp]
	\centering
	\includegraphics[width=\columnwidth]{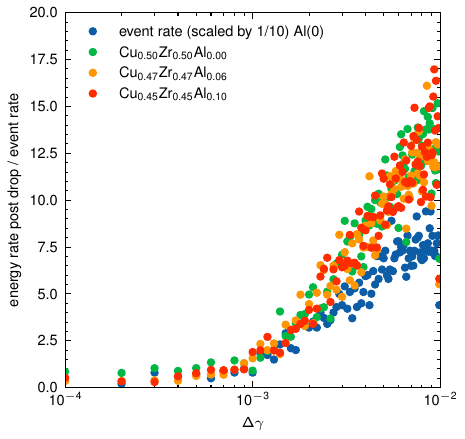}
	\caption{Post-stress drop correlations in the steady-state flow: event rate, and the stress drop rate for systems with different Al contents.}
	\label{figure5}
\end{figure}

Fig.~\ref{figure5} finally shows in the post-yield regime how the drop rate and "stress production rate" (binned sum of stress drops) scales with $\Delta \gamma$. For very small strain increments the rates are low and constant in agreement with the trivial assumption of what follows from the gap distribution in this limit. Beyond a strain of about 10$^{-3}$ both rates start to increase apparently linearly to approach the average event and stress drop rates.

\section{Conclusions}The modelling studies of metallic glasses have not tried much to handshake wih the general theory of glass mechanical properties. Here we have taken a look at the typical CuZrAl system under athermal quasistatic shear conditions regarding the early strain part, the typical mechnical properties and the steady-state flow beyond the yield point.

The impact of composition and cooling rate are found to be as expected as regards the impact on elastic modulus and yield stress. The development of plastic deformation shows some interesting differences as regards the impact of "how good the glass is" (cooling) but again adding more and more Aluminium shows that the composition has little effect in that all the species contribute to the deformation.

The statistics of stress drops exhibit some differences from previous studies. The early time response is typical of virgin disordered systems with a positive gap exponent. However, avalanches are not really common until strains of a substantial fraction of the yield strain are reached. The steady-state flow is found to be entertainingly superuniversal in that for all the (many) cases studied here we find the same exponent, roughly in line with earlier works, and the same cut-off. This prediction would be interesting to test in experiments.

\section{Acknowledgments}
\begin{acknowledgments}
T.M. and M.J.A. acknowledge the support from FinnCERES flagship (grant no. 151830423), Business Finland (grant nos. 211835, 211909, and 211989), and Future Makers programs. 
M.J.A. acknowledges support from the Academy of Finland Center of Excellence program (program nos. 278367 and 317464), as well as the Finnish Cultural Foundation. 
A.D.S.P, S.B. and M.J.A. acknowledge support from the European Union Horizon 2020 research and innovation program under Grant Agreement No. 857470 and the European Regional Development Fund via the Foundation for Polish Science International Research Agenda PLUS program under Grant No. MAB PLUS/2018/8. 
S.B. acknowledges support through SONATABIS grant DEC-2023/50/E/ST3/00569 from the National Science Center in Poland.
The authors acknowledge the computational resources provided by the Aalto University School of Science “Science-IT” project.
\end{acknowledgments}


\end{document}